\documentclass[conference]{IEEEtran}
\IEEEoverridecommandlockouts
\usepackage{cite}
\usepackage{amsmath,amssymb,amsfonts}
\usepackage{algorithmic}
\usepackage{graphicx}
\usepackage{textcomp}
\usepackage{xcolor}
\usepackage{longtable}
\usepackage{adjustbox}
\def\BibTeX{{\rm B\kern-.05em{\sc i\kern-.025em b}\kern-.08em
    T\kern-.1667em\lower.7ex\hbox{E}\kern-.125emX}}
\begin{document}

\title{COVID-19 Classification Using Deep Learning Two Stage Approach}

\author{\IEEEauthorblockN{1\textsuperscript{st} Mostapha Al Saidi}
\IEEEauthorblockA{\textit{EECS Department} \\
\textit{Florida Atlantic University}\\
Boca Raton, FL, USA \\
malsaidi2015@fau.edu}
\and
\IEEEauthorblockN{2\textsuperscript{nd} Ali Salem Altaher}
\IEEEauthorblockA{\textit{EECS Department} \\
\textit{Florida Atlantic University}\\
Boca Raton, FL, USA \\
aaltaher2018@fau.edu}
\and
\IEEEauthorblockN{3\textsuperscript{rd} Muhammed Tanveer Jan}
\IEEEauthorblockA{\textit{EECS Department} \\
\textit{Florida Atlantic University}\\
Boca Raton, FL, USA \\
mjan2021@fau.edu}
\and
\IEEEauthorblockN{4\textsuperscript{th} Ahmed Altaher}
\IEEEauthorblockA{\textit{EECS Department} \\
\textit{Florida Atlantic University}\\
Boca Raton, FL, USA \\
aaltaher2020@fau.edu}
\and
\IEEEauthorblockN{5\textsuperscript{th} Zahra Salekshahrezaee}
\IEEEauthorblockA{\textit{EECS Department} \\
\textit{Florida Atlantic University}\\
Boca Raton, FL, USA \\
zsalekshahre2018@fau.edu}
}

\maketitle

\begin{abstract}
In this paper, deep-learning-based approaches namely fine-tuning of pretrained convolutional neural networks (VGG16 and VGG19), and end-to-end training of a developed CNN model, have been used in order to classify X-Ray images into four different classes that include COVID-19, normal, opacity and pneumonia cases. A dataset containing more than 20,000 X-ray scans was retrieved from Kaggle and used in this experiment. A two stage classification approach was implemented to be compared to the one-shot classification approach. Our hypothesis was that a two stage model will be able to achieve better performance than a one-shot model. Our results show otherwise as VGG16 achieved 95\% accuracy using oneshot approach over five fold of training. Future work will focus on a more robust implementation of the two stage classification model Covid-TSC. Main improvement will be allowing data to flow from the output of stage1 to the input of stage2, where stage1 and stage2 models are VGG16 models fine-tuned on the Covid-19 Dataset.
\end{abstract}


\section{Introduction}
The novel 2019 Coronavirus infection (COVID-19) was found as a novel illness pneumonia within the city of Wuhan, China at the conclusion of 2019. Presently, it gets to be a Coronavirus episode around the world, the number of contaminated individuals and passings are expanding quickly each day concurring to the overhauled reports of the World Health Organization (WHO). \cite{i1}.

Due to its flu-like indicators and possibly genuine outcomes, a sensational increment of suspected COVID-19 cases are anticipated to overpower the healthcare system amid the flu season. Health system still generally distribute offices and resources such as Emergency Department (ED) and Intensive Care Unit (ICU) on a receptive way confronting critical labor and financial limitations. To optimize asset utilization and clinical workflow, a fast, mechanized, and point-of-care COVID-19 persistent administration innovation that can triage (COVID-19 case screening) and take after up (radiological trajectory prediction) patients is critically needed \cite{i2,i10}.

It is significant to utilize the screening strategies available to recognize COVID-19 cases and separate them from other conditions \cite{i3}. The authoritative test for COVID-19 is the Reverse Transcriptase - Polymerase Chain Reaction (RT-PCR) test \cite{i4}, which must be performed in specialized research facilities. COVID-19 patients, may have few interesting clinical and para-clinical highlights, e.g., showing anomalies in restorative chest imaging with commonly two-sided inclusion. The highlights were appeared to be discernible on chest X-ray (CXR) and CT scans \cite{i5,i11}. Since those images are tolerably characteristic to the human eye and not simple to recognize from pneumonia features, an AI-based methods have been utilized in various scenarios, counting robotized analyze and treatment in clinical settings. Deep Neural Networks (DNNs) have as of late been utilized for the conclusion of COVID-19 from therapeutic pictures, driving to promising results\cite{i5,i6,i7,i8,i17,i12}.

The rest of the paper is organized as following: Section.II presents related works done on the same topic. Section.III details our new approach (Covid-19-TSC) to solve this classification task and illustrates related experimental setup. Also, this section gives a brief explanation of other methods applied for the seek of comparison to our proposed model. Section.IV represents the results of the experiments applied. Section.V concludes and discusses the results.

\section{Related Works}

A (COVIDX-Net) system was proposed to consequently help the early diagnoses of patients with COVID-19 in an productive way utilizing the ordinary chest X-ray. COVIDX-Net underpins the improvement of procedures for CAD frameworks to battle the COVID-19 outbreak. Tests and assessment of the COVIDX-Net have done based on 80-20\% of X-ray pictures for the training and testing stages, separately. The VGG19 and the Densely Connected Convolutional Network (DenseNet) models showed a comparable execution of the automated COVID-19 classification with f1-scores of 0.89 and 0.91 for typical and COVID-19 cases, separately \cite{i1}.

A light weight Deep Neural Network (DNN) motivated mobile software, COVID-MobileXpert, for on-mobile COVID-19 quiet triage and take after up utilizing CXR pictures at the patient’s residents has been created. A three-player knowledge transfer and distillation (KTD) system counting a pre-trained  APN  that extricates CXR imaging highlights from a huge scale of lung infection CXR pictures, a fine-tuned RFN  that learns the fundamental CXR imaging highlights to segregate COVID-19 from pneumonia and/or ordinary cases with couple COVID-19 cases, and a prepared lightweight MSN to function on-device COVID-19 sufferer triage and follow-up. At the ED, COVID-MobileXpert calculates COVID- 19 probabilistic hazard to help mechanized triage of COVID-19 patients. At the ICU or general ward (GW), it employs a series of longitudinal CXR images to decide whether there's an looming weakening within the well-being condition of the COVID-19 sufferers \cite{i2}.Similarly COVID 19 diagnosis system using optimized and enhanced version of ECN using mayfly optimizations\cite{i20}

A ‘DeepCOVIDExplainer’ been proposed, which is a reasonable deep neural network (DNN)-based strategy for an automated classifying of COVID-19 side effects from CXR pictures. They utilized CXR pictures of sufferers,  healthy, pneumonia, and COVID-19 cases. CXR pictures are to begin with comprehensively pre-processing procedure, priory being grown and classified with a neural classifier, then, the areas are class-distinguished via the gradient-guided class activation maps (Grad-CAM++) and layer-wise relevance propagation (LRP) \cite{i13}.

A DarkCovidNet design, which is considered as a  deep learning structure was proposed for the automated classification of COVID-19. The proposed show has an end-to-end design without utilizing any feature extraction strategies, and it requires crude chest X-ray pictures to announce the classifications. The structure is trained with chest X-ray pictures, which are not in a normal shape and were gotten quickly. Symptomatic tests performed after 5–13 days are found to be positive in recuperated sufferers. This significant finding showed that the recouped sufferers may proceed to spread the virus. This strategy emphasized one of the foremost vital drawbacks of the chest radiography examinations is an inability to distinguish the early stages of COVID-19, as they don't have adequate affect-ability in GGO detection. Be that as it may, well-trained deep learning models can center on focuses that are not recognizable to the human eyes, and may serve to turn around this recognition \cite{i14}.

A GAN with deep transfer learning for coronavirus discovery in chest X-ray pictures was proposed. The need of datasets for COVID-19, particularly, in chest X-rays pictures was the most motivation of this method. The most thought was to gather all the conceivable pictures for COVID-19 that was existing and utilize the GAN network to produce more images to assist classifying this infection from the accessible X-rays pictures. The dataset utilized in this study was collected from diverse sources. There were four diverse sorts of classes, COVID-19, healthy, pneumonia bacterial, and pneumonia virus. Three deep transfer models were chosen, Alexnet, Googlenet, and Restnet18, since they contain a tiny number of layers in their structures, leading to diminished complexity, efficient memory and execution time of the proposed model. Three case scenarios were tested, in the first one, 4 classes classification, the Googlenet was chosen to be the core deep transfer model. In the second scenario, 3 classes classification, the Alexnet was selected. Whereas for the third situation, which included two classes (COVID-19, and ordinary), Googlenet was chosen \cite{i19}.

\section{Methodology}

This section details our approach in tackling the task of classifying X-ray images of four different classes. We compared the performance of various models on the classification of (Covid, Normal, Opacity, and Pneumonia) cases using X-rays images. We considered a comparative approach of classification between One-Stage classification and Two-Stage classification approaches. The models compared are a baseline CNN model that we constructed, VGG16 and VGG19. CNNs are being used in many fields as the baseline approach to any problem like emotion recognition\cite{i9}, person identification\cite{i18}, and object recognition in multiple domains such as industrial \cite{i22}, Medical \cite{i15, i21}, manufacturing and consumer side. The CNN model served as our baseline for comparison between CNN model learning from scratch versus pretrained high performing models such as VGG.

\subsection{Hypothesis}
The task at hand is a classification between normal and four different classes of viral infection. Due to high similarity in X-ray images and similarity between the features present in case of an infection (regardless of type) we hypothesized that a Two-Stage classification approach will increase accuracy and reduce the False-Negative classification rate. To test the hypothesis we designed an experimental procedure that produces the least variability in comparing the models. This was insured by using the same data files for training, validation and testing of our models. 

\section{Experiments}
The process of learning happens during training the model using the prepared data. After defining the model using the desired hyperparameters and depth of model, the process of training is initiated. In the case of CNN model, the initiation sets the weights to random values that will be updated with each epoch of training. 

Furthermore, the efficacy of the training process is evaluated using validation data. Validation data is acquired from the training dataset according to a 75/25 ratio split. Validation process happens in sync with training where after the model trains on the training data, validation data is passed and the accuracy and loss value are recorded at each epoch. Validating the learning process is an essential part of training deep learning models as it indicates the progress of learning and the validity of knowledge gained after each epoch. 

Finally to test the model a new set of data is passed to be classified post-training. This test is necessary as the model generates predictions based on data never seen before. This tests the generalization of the model, if the value was close to validation accuracy then the model generalizes well for X-Ray features. 

\subsection{Experimental Settings}

\subsubsection{One-Shot CNN}
The baseline approach is the typical one shot approach that we see in literature (cite few papers). The prepared data is used as an input to the model in order to be classified as [Normal, Covid, Opacity, Pneumonia] class. Our baseline model for all the approaches is the CNN model we constructed for this task.
\subsubsection{Covid-TSC}
\begin{figure}[htbp]
    \centerline{\includegraphics[width=0.5\textwidth]{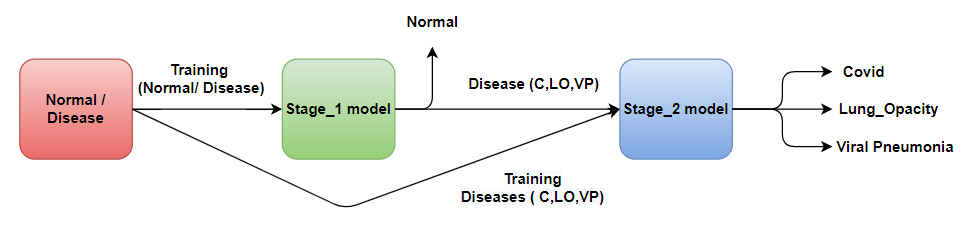}}
    \caption{Covid-TSC Model}
    \label{fig:tsc}
\end{figure}

Covid-TSC refer to our approach in solving this classification problem. Covid-TSC stands for Covid Two Stage Classification. The first stage of this approach is classifying the data between Disease class (Covid, Pneumonia, opacity) and Normal Class. For this stage, the data was prepared in away that combines equal distribution of disease classes and acquire an equal number of samples for the normal class. The models are then trained on this data to perform binary classification task. 
Following Stage1 we start Stage2 which is classification of the disease class itself into (Covid, Pneumonia, Opacity). 
As indicated in the hypothesis, our expectation is to get higher accuracy on the COVID classification.
A flowchart of the model is shown in Figure.\ref{fig:tsc}

\subsubsection{Technical Setup}
Our baseline model and transfer learning models (VGG) were implemented using Tensorflow with Keras backend library. This library is a specialized deep learning library that facilitates the construction, training and testing of deep learning models.
Furthermore, the experiment was coded using Python programming language as it is one of the most popularly used programming language for such applications. Finally the models were trained and tested using FAU HPC services that allow for GPU access. 

\subsection{Benchmark Data}
\begin{figure}[htbp]
    \centerline{\includegraphics[width=0.55\textwidth]{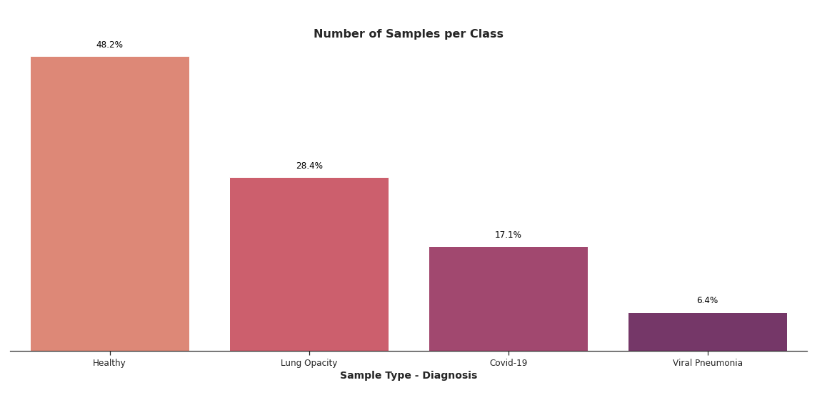}}
    \caption{Class Distribution}
    \label{fig:dist}
\end{figure}

Covid Radiography database retrieved from Kaggle contains an Image folder that contains four subfolders, one for each class. Since we are implementing a one-shot approach we keep the directory organization the same since we will be sampling data from each class as needed. Overall we queried 1500 X-rays from the directory for each class and stored each file label and path in a dataframe to be used for training. The 1500 X-rays used were further split to training and testing data following an 80/20 split, where 80\% of the data was used for training. Out of the training data, a further split was done following the 75/25 rule to split the data between training and validating sets. 
Furthermore, flow from the dataframe method from Keras was used to flow the data using an image data generator from the dataframe into the model. Label was given to each sample as one of the four classes known. The encoding of these labels is done by Keras through using ImageDataGenerator function. As we can see in \ref{fig:dist}, the dataset is quite imbalance and this is common phenomena when it comes to medical images. Many methods to overcome this problem proposed one such methods extraction features of from the current dataset using autoencoders and generate more data based on that to balance the dataset \cite{i20}

The database had 21,165 samples at the time of retrieval and the class distribution is shown in Figure.\ref{fig:dist}. 
Moreover, if we examine the image colour mean and image channel colour standard deviation we can visualize the distribution space of the different classes as shown in Figure.\ref{fig:class-scatter}. By examining the scatter plot we can see the normal class slightly overlaps other classes as compared to the overlap between the disease class samples.

\begin{figure}[htbp]
    \centerline{\includegraphics[width=0.55\textwidth]{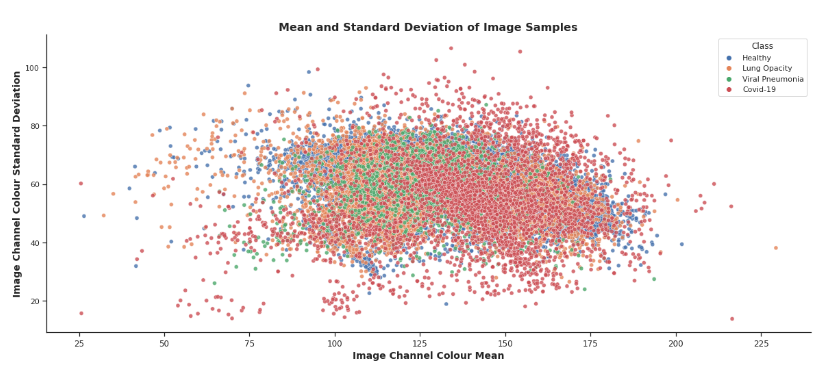}}
    \caption{Samples Scatter}
    \label{fig:scatter}
\end{figure}

Finally, we also examined the scatter plots of each individual class shown in Figure.\ref{fig:class-scatter} for differences and the findings were interesting. Specifically, Covid-19 scatter does not resemble any of the other three classes. It presents more outliers than the other classes, and the points are more scattered across the graph. Which could indicate that the images have a higher distinction between each other.

\begin{figure}[htbp]
    \centerline{\includegraphics[width=0.55\textwidth]{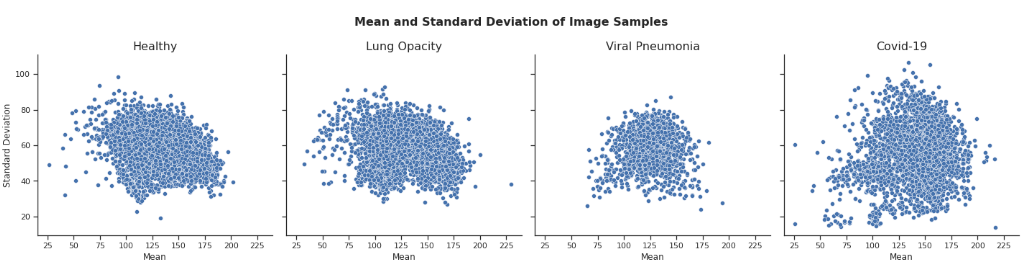}}
    \caption{Class Scatter}
    \label{fig:class-scatter}
\end{figure}

\subsection{Baseline Methods}
\subsubsection{Models}
In this experiment we compare the performance of three different deep learning model on the classification task of X-Ray scans between four different classes. Our base model is the CNN model constructed and trained from scratch. This model is considered as the baseline model and is compared to the performance of VGG16 and VGG19 that were fine-tuned using transfer learning approach. 
\subsubsection{Baseline Model-CNN}
\begin{figure}[htbp]
    \centerline{\includegraphics[width=0.5\textwidth]{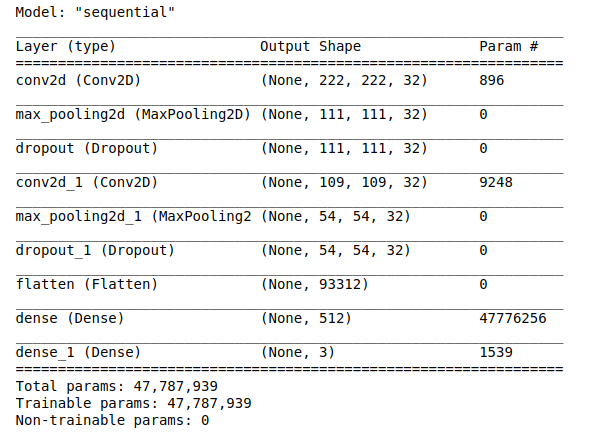}}
    \caption{CNN Model Summary}
    \label{fig:cnn}
\end{figure}

Convolutional Neural Network has had groundbreaking results over the past decade in a variety of fields especially computer vision related tasks, such as image classification. CNN’s revolutionized computer vision and led to the development of many high performance deep models such as VGG19.
The model summary shown in Figure.\ref{fig:cnn}. The model is considered to be very simple compared to VGG16 and VGG19 in Figures.\ref{fig:class-scatter}. The main block of any CNN model is the convolutional blocks. In our case, we have two Conv2D blocks and two Dense fully connected layers, leading the model to have 47,787,939 parameters to be trained. 
Further more, the activation function was set to 'Softmax' for multiclass classfication (one-shot approach) and 'Sigmoid' for Binary classification. Finally, the loss function to be minimized by the learning algorithm was set to "Categorical-CrossEntropy" for multiclass classification and "Binary-CrossEntropy" for binary classification.

\subsection{VGG16 and VGG19}

An alternative to constructing a CNN model and training it from scratch is to implement a state of the art deep model such as VGG16 and VGG19. These models are orders of magnitude bigger than our baseline model. Training such models from scratch is very computationally expensive, so a solution is to use a transfer learning approach and fine tune the model for a few epochs. 

\subsubsection{VGG Models}
\begin{figure}[htbp]
    \centerline{\includegraphics[width=0.5\textwidth]{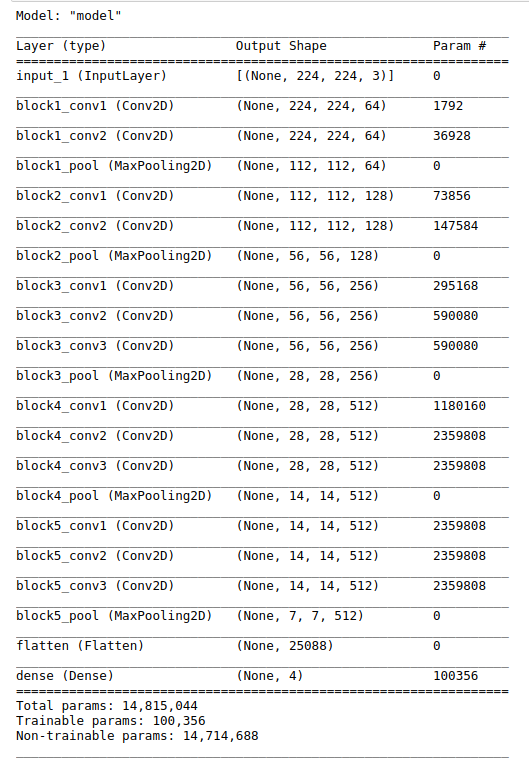}}
    \caption{CNN Model Summary}
    \label{fig:vgg}
\end{figure}

VGG16 and VGG19 (referred to as VGG) is a unique deep learning model that is based on CNN architecture. VGG is considered one of the best performing models architecture till date after winning the ILSVRC (Imagenet) competition in 2014. The model was implemented using Keras API with pre-trained imagenet weights. It is notable to mention that the VGG16 model has over 134 million parameters to train, whereas, VGG19 has over 138 million parameters. This is where transfer learning saves the day by reducing that number to 25,000 for both VGG16 and VGG19. The summary of these models is shown in Figure.\ref{fig:vgg}.

\subsection{EfficientNet}
EfficientNet \cite{i16} is a novel approach for scaling up CNNs that are more like organized and structured. Each dimension of the model is scaled in a uniform manner by passing a static parameters. With this method researchers were able to achieve 10x better efficiency more accurately its 8.4x smaller and about 6.1 x faster than the best CNNs available. That's not the only achievement, it is also able to achieve a 91.7\% accuracy on the CIFAR-100 and a whopping 98.8\% accuracy on the Flowers dataset.
\begin{figure}[htbp]
    \centerline{\includegraphics[width=0.5\textwidth]{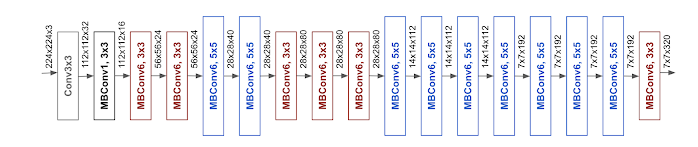}}
    \caption{EfficientNet-B0 Architecture}
    \label{fig:effb7}
\end{figure}

\subsection{Transfer Learning}
Transfer learning in summary is using stored knowledge gained from solving a problem to solve a different but similar problem. With transfer learning, instead of starting the learning process by training the model from scratch, we start from patterns that have been learned solving a different problem. We used state of the art deep learning such as VGG16 and VGG19. The models were pre-trained on the imagenet image library storing knowledge on visual patterns that are inherently present in almost all image datasets. 
\subsection{Results}
The models were trained using a 5-Fold cross validation technique to ensure that the learning is validated and tested. At the end of each epoch we collect the data into a csv file to be later used in plotting the Accuracy data for the models. 

\subsubsection{Accuracy and Loss Plots}
Figures 7 through 12 show the accuracy and loss plots for each model with three setups that are multi-class (four classes), Stage1 (Normal vs Disease) and Stage2 (Disease).

\begin{figure}[htbp]
    \centerline{\includegraphics[width=0.5\textwidth]{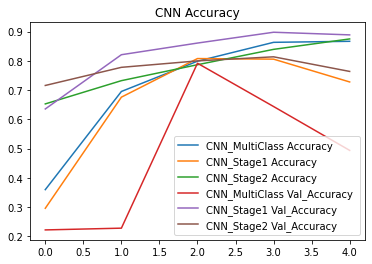}}
    \caption{}
    \label{fig:cacc}
\end{figure}

\begin{figure}[htbp]
    \centerline{\includegraphics[width=0.5\textwidth]{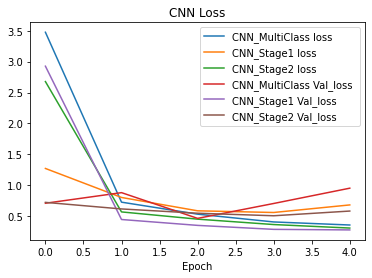}}
    \caption{}
    \label{fig:closs}
\end{figure}

\begin{figure}[htbp]
    \centerline{\includegraphics[width=0.5\textwidth]{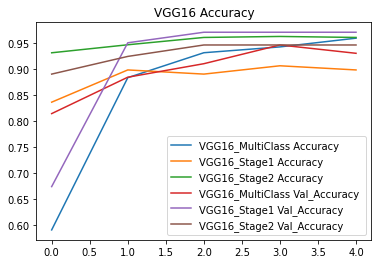}}
    \caption{}
    \label{fig:16acc}
\end{figure}

\begin{figure}[htbp]
    \centerline{\includegraphics[width=0.5\textwidth]{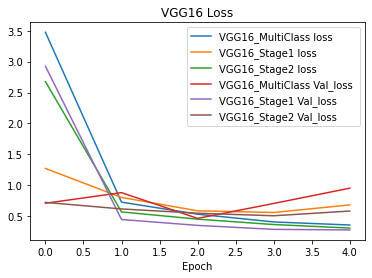}}
    \caption{}
    \label{fig:16loss}
\end{figure}

\begin{figure}[htbp]
    \centerline{\includegraphics[width=0.5\textwidth]{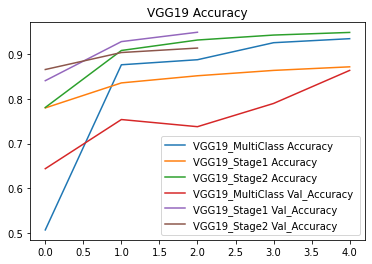}}
    \caption{}
    \label{fig:19acc}
\end{figure}

\begin{figure}[htbp]
    \centerline{\includegraphics[width=0.5\textwidth]{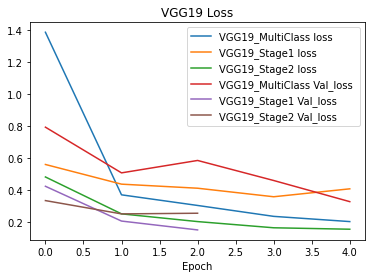}}
    \caption{}
    \label{fig:19loss}
\end{figure}

\subsection{Analysis of Results}

Although validation accuracy and loss combined with test accuracy are good metrics to indicate the performance of the model, they lack understanding of types of error occurring and misclassification.

	Classification report is used to measure the quality of predictions. More precisely, how many predictions are True and how many are False. More specifically, True Positives, False Positives, True negatives and False Negatives. These values are used to report metrics such as “Precision, Recall, F1-Score and support”. And these metrics are explained below. 

Precision is the ability of a classifier not to label an instance positive that is actually negative.
Recall is the ability of a classifier to find all positive instances.
F1 score is used to find the percentage of correct positive predictions. Mostly used for comparing models. 
Support is the number of actual occurrences of the class in the specified dataset.

Figure.\ref{fig:16acc} show that the model VGG16 achieved very high validation accuracy over the different stages of classifications, where the model achieved greater than 95\% on Stage1 classification. Furthermore, the model is also not overfitting as we can see form the accuracy and validation curves. In addition to the accuracy plots, the model also shows minimization of the loss function. 
Since VGG16 model appears to be superior to both CNN and VGG19 we are going to proceed with this model. 
It is important to note that although VGG19 is deeper than VGG16 as it has three additional convolutional layers, it does not outperform VGG16, proving that the depth of the model does not guarantee superiority of performance. 

\subsubsection{VGG16 Confusion Matrix}
Confusion matrix shows a lot of important information about the performance of the model as it displays the number of correct predictions and misclassifications for each class. 
F
\begin{figure}[htbp]
    \centerline{\includegraphics[width=0.5\textwidth]{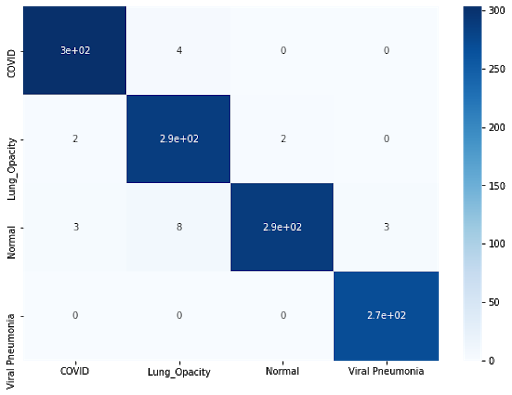}}
    \caption{}
    \label{fig:multi}
\end{figure}

\begin{figure}[htbp]
    \centerline{\includegraphics[width=0.5\textwidth]{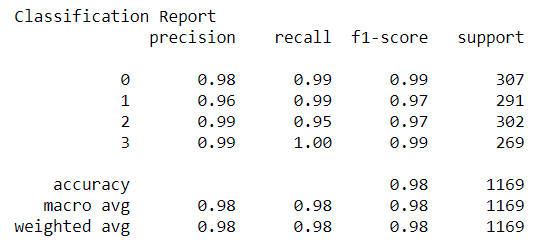}}
    \caption{Multiclass Confusion Matrix }
    \label{fig:cr_multi}
\end{figure}

Figure.\ref{fig:multi} shows the confusion matrix generated from training VGG16 over the four classes of data. 
In addition to the confusion matrix, classification report was generated using Sklearn library and shown in Figure.\ref{fig:cr_multi}. The report shows Covid Recall value to be 0.99 which is extremely important. It means that the model is able to correctly identify 99\% of Covid cases. Furthermore, precision is 0.98 which means that the model might label 2\% of non-Covid images as Covid. 

\begin{figure}[htbp]
    \centerline{\includegraphics[width=0.4\textwidth]{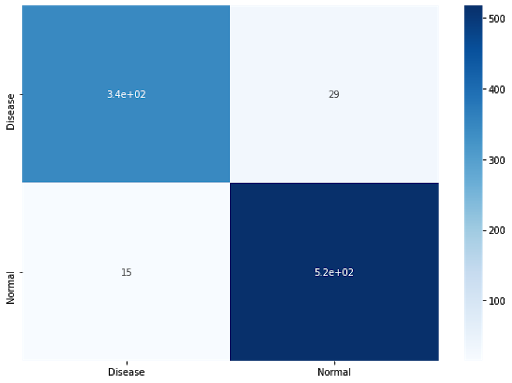}}
    \caption{Stage1 Confusion Matrix }
    \label{fig:nd}
\end{figure}

\begin{figure}[htbp]
    \centerline{\includegraphics[width=0.4\textwidth]{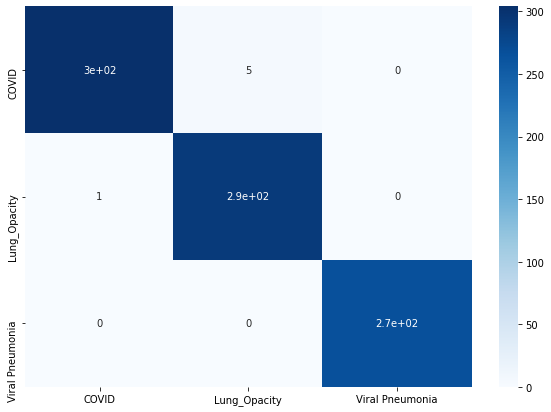}}
    \caption{Stage2 Confusion Matrix}
    \label{fig:d}
\end{figure}

Figures.\ref{fig:nd} and Figure.\ref{fig:d} show the confusion matrix for Stage1 and Stage2 classification respectively. And Figure.\ref{fig:nd} and Figure.\ref{fig:d} display the Classification report over Stage1 and Stage2 respectively. 
Covid19-TSC results are shown in figures 15-18. Figure \ref{fig:d} shows the results of Stage2 which is our main model to inspect as it classifies the disease data to their respective classes. Figure.18 shows covid class has 0.98 recall whereas opacity and pneumonia has 1.00 recall. Which means the model is able to correctly identify opacity and pneumonia cases 100\% of the time and covid cases 98\% of the time. Although the performance is highly reliable, we cannot accept that this experiment proves our hypothesis and further work is need to enhance this Covid19-TSC model.
\begin{figure}[htbp]
    \centerline{\includegraphics[width=0.4\textwidth]{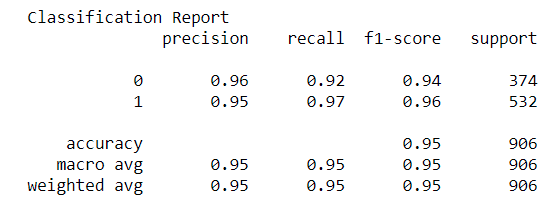}}
    \caption{Stage1 Classification Report}
    \label{fig:nd}
\end{figure}

\begin{figure}[htbp]
    \centerline{\includegraphics[width=0.4\textwidth]{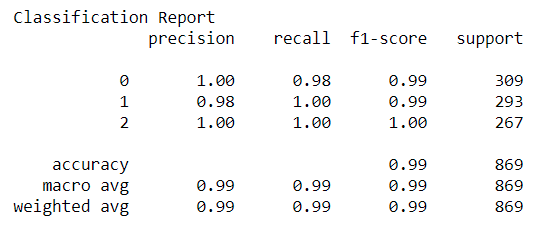}}
    \caption{Stage2 Classification Report}
    \label{fig:d}
\end{figure}

\subsection{Resource Comparison}
Resource comparison is mostly based on the hardware use. In order to get accurate and consistent results we kept using the same hardware resources which are mentioned below.
CPU : AMD Ryzen 5 5600x \newline
RAM : T-Force 16GB 3200mhz \newline
GPU : Nvidia GeForce RTX 3070 \newline

\begin{table}
\begin{tabular}{ |p{2cm}||p{2.5cm}|p{1cm}|p{1.5cm}|  }
 \hline
 \multicolumn{4}{|c|}{Resources} \\
 \hline
 Method/Model &  Dataset & Time & Accuracy\\
 \hline
VGG16 & Covid Radiology & 3.14s & 84.8\\
VGG19 & Covid Radiology & 3:50s & 83.8 \\
EfficientNet & Covid Radiology & 12:16s & 84.9 \\
Covid-TSC & Covid Radiology & 2.10 & 82.4 \\
 \hline
\end{tabular}
\caption{Resources Utilized}
\label{table:1}
\end{table}

Above table shows a detail comparison of the models used in this paper and its related attributes.Table \ref{table:1} is and overview of the models that were trained on the COVID-Radiology dataset using transfer learning for VGG16, VGG19 and EfficientNet. It also show the amount time each model took to train and classify the data of given dataset with the system mentioned above. Although the table \ref{table:1} shows that our model achieved less accuracy then other models but time needed to train the model was less then all of them.

\begin{table}
\begin{tabular}{ |p{2cm}||p{2.5cm}|p{1cm}|p{1.5cm}|  }
 \hline
 \multicolumn{4}{|c|}{Comparative Analysis} \\
 \hline
 Model & Dataset & Parameters & Accuracy\\
 \hline

VGG16 & ImageNet (14.2m) & 138m & 95.2\\
VGG19 & ImageNet (14.2m)& 143m & 91.3 \\
EfficientNet & ImageNet (14.2m) & 64m & 84.90 \\
Covid-TSC & Covid Rad. (23k) & 8m & 82.4 \\

\hline
\end{tabular}
\caption{Models Attributes}
\label{table:2}
\end{table}

In table \ref{table:2} we shows a comparative analysis of why our model was able to train in less time and still able to achieve a fair metrics as compared to other models. COVID-TSC as compared to Other models are comprised of a huge number of parameters and are also trained on very large dataset which in this case ImageNet. The difference between the two datasets are tremendous and same is case with number of parameters which are almost 9x for EfficientNet and 17x for VGG models while COVID-TSC was trained on COVID Radiology dataset and number of parameters for that was only 8 millions.

\section{Results}

\section{Conclusion}

In this project we trained three different deep learning models in order to classify four classes of X-Ray scans. The data was related to Covid, opacity, Pneumonia, and Normal cases. The models trained were based of Convolutional Neural Networks with our baseline model being CNN and two state of the art deep learning models VGG16 and VGG19. 

Furthermore, we wanted to test our hypothesis of dissecting the classification task to two stages were the first stage is distinguishing normal from disease class and the second stage is distinguishing which disease it is. 

Upon training and testing our models, we found that using transfer learning and VGG16 model was superior to all the other methods. VGG16 achieved greater than 95\% on testing accuracy over five folds of training. 

VGG16 model was superior to the other models in all the proposed approaches of training. However, this experiment proved our hypothesis to be invalid. VGG16 achieved less accuracy and Recall on the two stage approach versus the one shot approach. 
Future work will focus on resetting our experimental setup to allow for data to flow from the output of Stage1 to the Input of Stage2. We think that this approach will optimise feature extractions for the disease data. Additionally, further fine-tuning is need and hyper-parameter optimization. 

In conclusion, in this project we utilized Keras pretrained VGG16 and VGG19 models and implemented transfer learning approach to achieve our task. We also trained a CNN model that is able to achieve expert level predictions with few trainable parameters compared to VGG.

\end{document}